# Superconducting and structural properties of non-centrosymmetric Re$_6$Hf superconductor under high pressure

Sathiskumar Mariappan[1,2], Manikandan Krishnan[1], Dilip Bhoi[2], Hanming Ma[2], Jun Gouchi[2], R. P. Singh[3], Kapil Motla[3], Ponniah Vajeeston[4], Arumugam Sonachalam[1*] and Yoshiya Uwatoko[2*]

[1]Centre for High Pressure Research, School of Physics, Bharathidasan University, Tiruchirappalli, 620024, India.
[2]Institute for Solid State Physics, The University of Tokyo, 5-1-5 Kashiwanoha, Kashiwa, Chiba, 277-8581, Japan.
[3]Indian Institute of Science Education and Research Bhopal, Bhopal, 462066, India.
[4]Department of Chemistry, Centre for Materials Science and Nanotechnology, University of Oslo, P.O Box 11126, Blindern, N-0318 Oslo, Norway.
[*]**Corresponding authors:** sarumugam1963@yahoo.com, uwatoko@issp.u-tokyo.ac.jp

## Abstract

We report the effect of high pressure on the superconducting, vortex pinning, and structural properties of a polycrystalline non-centrosymmetric superconductor Re$_6$Hf. The superconducting transition temperature, $T_c$, reveals a modest decrease as pressure ($P$) increases with a slope -0.046 K/GPa (-0.065 K/GPa) estimated from resistivity measurements up to 8 GPa (magnetization measurement ~ 1.1 GPa). Structural analysis up to ~18 GPa reveals monotonic decreases of lattice constant without undergoing any structural transition and a high value of bulk modulus $B_0 \approx 333.63$ GPa, indicating the stability of the structure. Furthermore, the upper critical field and lower critical field at absolute temperature $(H_{c2}(0) \& H_{c1}(0))$ decreases slightly from the ambient pressure value as pressure increases up to 2.5 GPa. In addition, analysis of $\rho(T, H)$ up to $P \sim 2.5$ GPa using thermally activated flux flow of vortices revealed a double linearity field dependence of activation energy of vortices ($U_o(H)$), confirming the coexistence of single and collective pinning vortex states. Moreover, analysis of critical current density using the collective pinning theory showed the transformation of $\delta T_c$ to $\delta l$ pinning as pressure increases, possibly due to migration of grain boundaries. Besides, the band structure calculations using density functional theory show that density of states decreases modestly with pressure, which may be a possible reason for such a small decrease in $T_c$ by pressure.

**Introduction**

The discovery of superconductivity in non-centrosymmetric (NCS) compounds, which lack inversion symmetry, has encouraged theoretical and experimental investigations because these compounds can host unconventional superconductivity. For example, magnetic NCS superconductor $CePt_3Si$ has shown the signature of mixed parity states involving spin-singlet and triplet cooper pairs in the superconducting state [1,2]. However, the superconducting pairing glue forming the cooper pair in these NCS, such as $CePt_3Si$ [1,3], $Ce(Rh,Ir)Si_3$ [4,5], $Ce(Co,Ir)Ge_3$ [6,7] is complicated by the fact that the coexistence of magnetic order and superconductivity. Moreover, when such heavy fermionic NCS is pressurized, the long range magnetic order is continuously suppressed and superconductivity appears in a wide $P$ range [5,8–13]. Similarly, in both $CePdSi_2$ and $CeIn_3$ antiferromagnetic order decreased monotonically, followed by the appearance of superconductivity above critical pressure ($P_c$) [14]. Unlike these magnetic compounds, numerous non-magnetic NCS superconducting compounds, such as $LaNiC_2$ ($T_c \sim 2.8\ K$) [15,16], $Li_2(Pd,Pt)_3B$ ($T_c \sim 7\ K\ \&\ 2.7\ K$) [17,18], $(Ta,Nb)Rh_2B_2$ ($T_c \sim 6\ K\ \&\ 7.6\ K$) [19–21], $BiPd$ ($T_c \sim 3.8\ K$) [22,23], $SrPtAs$ ($T_c \sim 2.4\ K \sim 2.4$ K) [24,25], have also been studied and found to exhibits unconventional nature of superconductivity.

Apart from these compounds, in recent years, we have seen a quick development in both experimental and theoretical studies on new Re-based non-magnetic NCS compounds such as $Re_6(Zr,Ti)$ ($T_c \sim 6.75\ K\ \&\ 6\ K$) [26–28], $Re_3Ta$ ($T_c \sim 4.68\ K$) [29], $Re_{24}Ti_5$ ($T_c \sim 6\ K$) [30,31], $Re_{0.82}Nb_{0.18}$ ($T_c \sim 8.8\ K$) [32,33]. The $H_{c2}(0)$ of these NCS is very close to the Pauli paramagnetic limit, $H_p \sim 1.86 \times T_c$. Besides, the temperature dependence of heat capacity below $T_c$ is well explained by the s-wave symmetry of the cooper pairs [28,30]. Although these results are consistent with the expectations for a conventional BCS type superconductor, however, muon spin rotation experiments have revealed the signature of time reversal symmetry breaking in $Re_6X$ (X = Zr, Ti, Hf) [26,27,34]. Furthermore, $Re_6Hf$ has shown type-II superconductivity with the signature of possible mixing of spin singlet and triplet states [34].

Besides, the investigation of superconducting properties at ambient pressure, there are a few reports on the effects of chemical pressure as well as external pressure on the superconducting properties of $Re_6Hf$ [35]. On substituting Fe in place of Re, $T_c$ of $Re_6Hf$ is suppressed rapidly suggesting that the superconductivity is drastically affected by the doping of magnetic impurities. In contrast, by the application of external $P$ up to 2.5 GPa, the $T_c$ of Re$_6$Hf changes hardly [35]. Very recently, good quality polycrystalline samples of Re$_6$Hf have been synthesized via the arc-melting technique, and the structural and superconducting properties have been reported [36]. In this work, using high quality polycrystalline Re$_6$Hf samples, we have studied the effect of pressure on the structural and electronic properties of Re$_6$Hf via transport (~8 GPa), magnetic (~1 GPa) and powder X-Ray diffraction (~18.01 GPa) measurements. In addition, complementary density functional theory (DFT) calculations (up to ~ 25 GPa) have also been performed in order to understand the mechanism of superconductivity.

**Experimental techniques**

High quality polycrystalline Re$_6$Hf compound is prepared via arc-melt technique as described in the reference [36]. The details of phase purity, and sample characterization at ambient pressure were reported in Ref. [36]. The temperature dependent electrical resistivity ($\rho(T)$) was measured at ambient pressure with high magnetic field using oxford 18 Tesla magnet and AC resistance bridge. Also, $\rho(T)$ were measured by a standard four-probe method using the Physical Properties Measurement System (PPMS). $\rho(T)$ with the presence of a magnetic field ($H$) were measured at ambient and high pressure up to ~ 2.5 GPa using a double layer self clamp-type piston pressure cell which is made of Be-Cu and Ni-Cr-Al alloys. The pressure transmitting medium was Daphne Oil 7474 for its adequate hydrostatic nature to measure the electrical resistivity of synthesized compounds. The actual pressure inside the pressure cell was determined using the superconducting $T_c$ of a Pb specimen. The cubic anvil-type device was used to measure $\rho(T)$ for $P$ 2 to 8 GPa at the Institute of Solid State Physics, University of Tokyo, Japan. It consists of six tungsten carbide (WC) anvils to generate homogeneous hydrostatic pressure on the sample. The sample was placed in the Teflon capsule which was filled with the pressure medium (Daphne 7474) and then the Teflon cell was fixed in the pyrophyllite gasket and the gasket was placed between the six WC anvils. The electrical contacts were made using gold wire of 20 μm $\phi$ and Ag paste. The applied pressure was calibrated using resistive transitions of Bi I-II (2.55GPa), Bi II-III (2.7GPa) and III-IV (7.7GPa) at room temperature [37]. The temperature and field dependent magnetization measurements were measured using a magnetic property measurement system (MPMS-XL, Quantum Design, USA) at ambient and high $P$. The rectangular shaped sample with the dimensions, length ($l$) ≈ 0.281 cm, width ($b$) ≈ 0.105 cm and thickness ($t$) = 0.042 cm was used for all magnetic measurements. The magnetic field was applied parallel to the length of the sample. External applied $P$ up to ~ 1 GPa was produced by a Clamp-type micro pressure cell which is made of nonmagnetic Cu–Be alloy. The Daphne 7474 was used as a pressure transmitting medium and the actual pressure was calibrated from the $P$ dependence of superconducting $T_c$ of Sn. The high pressure X- ray diffraction pattern were recorded by a Rigaku, XtaLab HyPixsel 6000 diffractometer with Mo$K\alpha$ radiation ($\lambda = 0.71073$ Å) at room temperature used for obtaining the diffraction pattern of a single crystal. The obtained data were converted into intensity vs $2\theta$ pattern using the CrysAlis pro software. Pressure was generated using a Diamond Anvil Cell (DAC) with rhenium as a gasket (thickness 70 $\mu$m, hole $\phi$=0.1 mm). The mixture of methanol and ethanol (1:1) were used as a pressure transmitting medium, and the pressure was calibrated from ruby by the luminescence method. The emission signals ($R_1$ & $R_2$) of the ruby sample are received from other side of the cell through the gasket hole.

**Computational details**

The computations were carried out in the context of periodic density functional theory, as implemented in the VASP code [38] and examined Re$_6$Hf, which exhibits $I\bar{4}3m$ space group. The interaction of the core (Re:[Xe] and Hf: [Xe]) and valence electrons has been characterized using the projector-augmented wave (PAW) approach [39,40]. For the structural optimization, we used the Perdew, Burke, and Ernzerhof (PBE) gradient corrected functional for the exchange-correlation part of the potential. Our earlier [41] computations indicated that the only way to forecast structural characteristics in superconducting materials properly is to use a strong energy cutoff to ensure basis-set completeness. As a result, we chose a cut-off of 600 eV. When all atomic forces were less than 0.02 eV Å$^{-1}$, it was presumed that the atoms were relaxed, and the geometries were optimized when the total energy converged to less than 1 meV between two successive geometric optimization steps. The electronic characteristics of the structures optimized at the PBE level were computed. The following section discusses the reliability of this computational approach. We employed a Monkhorst–Pack $9 \times 9 \times 9$ k-mesh for structural optimization and electrical structure research unless otherwise noted. The band structures of the irreducible part of the first Brillouin zone were determined by solving the periodic Kohn–Sham equation at ten k-points along each direction of high symmetry. Total energy as a function of volume has been computed and fitted using the universal equation of state (EOS) [42]. Pressure vs. Gibbs free energy curves are used to compute the transition pressures. Gibbs free energy ($G = U + PV - TS$; $G$ = total energy + pressure × volume) is determined as follows: The volume vs total energy curve is fitted to the universal EOS function $P = \left(\frac{B_0}{B_0'}\right) \times \left[\left(\frac{v_e}{v}\right)^{B_0'} - 1\right]$. The relationship can be inverted to obtain the volume $(v) = v_e / \left[\left(1 + \left(\frac{B_0'}{B_0}\right) \times P\right)^{\frac{1}{B_0'}}\right]$. Where, respectively, $v_e, B_0,$ and $B_0'$ denote the equilibrium volume, the bulk modulus, and its pressure derivative. Following that, the bisection method is used to calculate the inverse. The enthalpy difference between the two data sets was computed using the scan over the pressures. According to the fitted energy-volume curve, the computed $B_0$ is 270 GPa and the $B_0'$ is 3.83.

Re$_6$Hf has approximately eight formula units per unit cell and is fundamentally disordered. Hf inhabited both 2a (0,0,0) and 8c (0.3106,0.3106,0.3106) sites, while Re occupied two 24g (0.3623,0.3623,0.0383) and 24g (0.6954,0.6954,0.2903) sites [36]. The composition of the fully stoichiometric phase is Re$_{48}$Hf$_{10,}$ with Hf occupancy of 80%. The ab initio random searching structure (AIRSS) [43] approach was used to generate alternative model structures for the specified substitution in the Re$_{48}$Hf$_{10}$ matrix, in addition to VASP calculations, for the partial occupancy simulation. As a result, Hf has been substituted for the vacancy in the acquired AIRSS model, and the structure has been completely relaxed. In this scenario, the composition of the theoretical sample (Re$_6$Hf) is almost identical to the composition of the experimental sample.

**Results and discussion**

Figure 1(a) shows the $\rho(T)$, of Re$_6$Hf under various applied pressure $(P)$ up to ~ 8 GPa. The sharp superconducting transition and zero resistivity at each pressure indicate the hydrostatic pressure is maintained during the whole set of measurements. At ambient pressure, as temperature decreases $\rho$ decreases from 163 μΩ cm at 300 K to 143 μΩ cm at 7 K (as shown in figure 1(b)). The residual resistivity ratio, defined as $\rho_{300 K}/\rho_{7 K} \sim 1.14$, is rather small due to the disorder dominated in this system and is similar to other Re-based NCS compounds [29,31,33,36]. With the application of pressure, the overall $\rho(T)$ slightly decreases. As $P$ increases from 0 to 8 GPa, $\rho_{300 K}$ decreases linearly with a slope of -0.11 μΩ-cm/GPa ($0 \leq P \leq 2.5$ GPa) thereafter follows a faster rate of decrease ~0.63 μΩ-cm/GPa. To further understand the normal state properties, we analyzed the $\rho(T)$ in the temperature range $10 K \leq T \leq 50 K$ at various $P$ using the Fermi-liquid relation, $\rho = \rho_0 + AT^2$, where $\rho_0$ and $A$ are residual resistivity and electron-electron scattering factor, respectively. As shown in figure 1(c), $\rho_0$ decreases from 149.18 μΩ-cm at ambient pressure to 140.75 μΩ-cm at 8 GPa. Simultaneously, $A$ decreases from 1.25 $\mu\Omega$-cm/K$^2$ at ambient pressure to 0.60 x10$^{-3}$ $\mu\Omega$-cm/K$^2$ at 8 GPa. The small value of $A$ suggests that Re$_6$Hf is a weakly correlated system.

To see the evolution of superconductivity under pressure, we plotted the $\rho(T)$ near the superconducting transition region at various pressures in Figure 2(a). The $\rho(T)$ transition shifts parallel towards low temperature with pressure. We estimated the onset of the superconducting transition temperature, $T_c^{on}$, from the intersection of two extrapolated lines. One is drawn through the normal state $\rho(T)$ curve and the other is drawn through the superconducting transition region as shown in figure 2(a). At ambient pressure, $T_c^{on} \sim 5.96$ K and the zero-resistivity temperature, $T_c^0 \sim 5.56$ K. The superconducting transition width, $\Delta T_c = T_c^{on} - T_c^0 \approx 0.40\ K$ at 0 GPa is rather sharp and comparable to a previous report [36]. With pressure, $T_c^{on}$ decreases to 5.59 K at 8.0 GPa and $\Delta T_c \sim 0.22\ K$ becomes sharper. Figure 2(b) shows temperature dependent zero-field cooled (ZFC) and field cooled (FC) magnetic susceptibility ($4\pi\chi$), measured with an applied magnetic field 2 mT from to $\sim$ 1.2 GPa. The estimated $4\pi\chi$ at 2 K reaches -0.96 suggesting the full Meissner state at ambient pressure. The $4\pi\chi$ decreases to -0.77 at 1.22 GPa indicating the reduction of Meissner fraction with increasing $P$. The ZFC magnetic curve displays a transition at $T_c^{on} \sim 5.95$ K at 0 GPa and 5.88 K at 1.2 GPa, which is in good agreement with the $T_c$ determined from $\rho(T)$ for $P < 1.5$ GPa. In Figure 2(c), using the magnetization (up to $\sim$ 1.2 GPa) and resistivity (up to $\sim$ 8 GPa), we summarized the $T_c$ vs $P$ curve for Re$_6$Hf, revealing a monotonic decrease of $T_c$ with $P$. The rate of suppression of $T_c$ with respect to the applied $P$, $dT_c/dP$, determined from $\rho(T, P)$ is $\sim$ -0.046 K/GPa whereas from $\chi(T)$ shows a slightly faster decrease, $dT_c/dP \sim -0.065\ K/GPa$ .

Figure 3(a)-(d) shows the magnetic field dependence of isothermal magnetization, $M(H)$, at various temperatures with different fixed applied $P$ of 0 GPa, 0.35 GPa, 0.81 GPa, and 1.22 GPa, respectively. In both ambient and high $P$, a linear response of the Meissner signal can be clearly seen in the $M(H)$ curve in the low magnetic field region. From the $M(H)$ curves, we determined the temperature depence of lower critical field ($H_{c1}(T)$) as the field where the Meissner signal deviates from linearity, as shown in figure 3(a). The deflection observed in isothermal magnetization curves are more noticeable and its slope is given by, $M = \frac{-H_\alpha}{1-N}$, where $H_\alpha$ is the external applied field and $N$ is the demagnetization factor. For an ellipsoidal sample, $N$ is given by the relation, $N = \left(\frac{2}{\pi}\right) arc \sin(1/(1 + 2(l/b)))$ , with $l$ and $b$ as the length and width of the samples, respectively [44]. We also calculated the $N$ assuming a prismatic shaped sample based on the relation $N = bc/(ab + bc + ca)$ [45]. For both methods, we obtained rather a small value of $N$ (~0.07 for an ellipsoidal sample and ~0.09 for a prismatic shaped sample). These small values of $N$ suggest that the demagnetization factor will not severely affect to the overall analysis. The obtained $H_{c1}(T)$ fitted with the quadratic Ginzburg-Landau (GL) relation [46], $H_{c1}(T) = H_{c1}(0)\left(1 - \left(\frac{T}{T_c}\right)^2\right)$ and is shown in figure 3(e). From the GL fitting, $H_{c1}(0)$, is estimated to be 5.88 mT at 0 GPa and found to decrease to 5.52 mT at 1.22 GPa as shown in figure 3(f).

To determine the $H_{c2}(0)$, we measured the $\rho(T)$ by fixing the magnetic fields [shown in figure 4((a)-(i))] as well as the field dependence of resistivity ($\rho(H)$) by fixing the temperature across the superconducting region at ambient and high $P$ [shown in figure 4 ((f)-(i)]. The temperature dependence of upper critical field ($H_{c2}(T)$)curve at different pressures is estimated as the magnetic field dependence of $T_c^{on}$ determined from the ($\rho(T)$ and $\rho(H)$ data). At ambient $P$, $\rho(H)$ is found to be zero even up to ~10 T above, which changes from superconducting to normal state. In figure 5, we plotted the $H_{c2}(T)$ at different pressures. To understand the superconducting pairing mechanism, we analyzed the $H_{c2}(T)$ using the single band Werthamer-Helfand-Hohenberg (WHH) relation, $H_{c2}^{orb}(0) = 0.693 T_c \left(\frac{dH_{c2}}{dT}\right)_{T_c}$ [47], where $\left(\frac{dH_{c2}}{dT}\right)_{T_c}$ is the slope of the $H_{c2}(T)$ curve at T = $T_c$ and $H_{c2}^{orb}(0)$ is the orbital pair breaking critical field at absolute zero. $P$ dependent of $H_{c2}^{orb}(0)$ is found to decrease from 9.78 T at ambient condition to 9.18 T at 2.5 GPa[figure 6(a)]. We also estimated the $H_{c2}(0)$ as shown in figure 5 using the Ginzburg-Landau (GL) theory, $H_{c2}(T) = H_{c2}^{GL}(0)\left(\frac{(1-t^2)}{(1+t^2)}\right)$, where $t$ is the normalized temperature $T/T_c$ and $H_{c2}^{GL}(0)$ is the upper critical field at zero temperature. At ambient pressure and high $P$, we found that $H_{c2}^{GL}(0)$ evaluated by fitting the GL formula is higher than the $H_{c2}^{orb}(0)$ [figure 6(a)]. From figure 6(a), it is also clear that, $H_{c2}^{GL}(0)$ is only slightly higher than the Pauli limited upper critical field, $H_p(0)$, while $H_{c2}^{orb}(0)$ is slightly lower than the $H_p(0)$. For weakly coupled BCS superconductors [48–50], $H_p(0) = 1.86 \times T_c$, which changes very weakly with $P$. The comparable values of the upper critical fields $H_{c2}^{orb}(0)$and $H_{c2}^{GL}(0)$, with that of $H_p(0)$ do not firmly support the existence of spin triplet paring in Re$_6$Hf. But the observation of time reversal symmetry breaking in recent $\mu sR$ study does not completely ruled out the possibilities of any spin triplet pairing in Re$_6$Hf contributing to the superconductivity [34]. Similar comparable values of the upper critical fields, $\mu_0 H_{c2}^{orb}(0)$ and $\mu_0 H_{c2}^{GL}(0)$ to the $H_p(0)$ and their $P$ dependence have also been observed in NCS TaRh$_2$B$_2$, which has also been investigated as a potential candidate with a spin triplet pairing [19]. Furthermore, the estimated Maki parameter [51], $\alpha = \sqrt{2}H_{c2}^{orb}(0)/H_p(0) \sim 1.24$ - 1.2 is higher than 1 [figure 6(b)], indicating the significant paramagnetic pair breaking effect. The superconducting GL coherence length relation is $H_{c2}^{GL}(0) = \Phi_0/2\pi\xi^2(0)$, where $\Phi_0$ is flux quantum (2.07x10$^{-15}$ Wb) and $P$ dependent $\xi(0)$ is shown in figure 6(c).

Upon application of a magnetic field, the resistivity in the superconducting transition region broadens and shifts towards a lower temperature due to dissipative thermally activated flux flow (TAFF) of vortices. In the TAFF regime, a finite resistivity occurs from the competition between two mechanisms: Flux creep (pinning force of vortices) and flux flow (Lorentz force acting on the vortices) [52]. In this region, resistivity follows the Arrhenius expression, $\ln(T, H) = ln\rho_0(H) - \frac{U_0(H)}{T}$, where $ln\rho_0 = ln\rho_{0f} + U_0/T_c$ and $U_0$ is the apparent activation energy which plays the crucial role as effective pinning barrier of vortices [53]. Both at ambient and high $P$, on plotting the $\ln(\rho(T,H))$ in the TAFF region as function of $1/T$, reveals a linear curve, thus confirming the Arrhenius behavior of resistivity in TAFF regime. The temperature dependences of thermally activated energy (TAE), U(T,H) reveals almost linear as observed from $U = U_0(H)\left(1 - \frac{T}{T_c}\right)$. On extrapolation, the fitted linear lines to $ln\ \rho(T,H)\ vs\ 1/T$ curve at various magnetic field, intersect at a temperature close to superconducting $T_c$. Moreover, in figure 7(a) $log\ \rho_0(H)\ vs\ U_0(H)$ shows a linear behavior and can be fitted with the relation $ln\rho_0 = ln\rho_{0f} + \frac{U_0}{T_c}$, the extracted parameters $\rho_{0f}$ and $T_c$ at both ambient and high $P$ are shown in figure 7(b). The linear relationship between the $(U_0)$ and $ln\rho_0$ for all pressures confirms the validity of our analysis. On increasing $P$ to 2.5 GPa, the values of $\rho_{0f}$ decreases to 5.16 m$\Omega$-cm from 12.96 m$\Omega$-cm at ambient pressure. $T_c$ also decreases to 5.16 K from 5.26 K at ambient pressure, in consistent with experimental behavior of $T_c$ under $P$, even though the estimated $T_c$ values are slightly lower.

$U_0(H)$ can be extracted from the Arrhenius relation, as $U_0(H) = \partial \ln\rho/\partial\left(\frac{1}{T}\right)$. In figure 7(a), we plotted the $U_0(H)$ at different pressures. From the figure 7(c), it is clear that the $U_0$ decreases both with the $H$ Field dependence of $U_0$ reveals two different linear regions suggesting the power law dependence of $U_0(H) \propto H^{-\alpha}$ and confirms the transition from a single vortex to collective (mixed) vortex state with an increasing field. The low magnetic field region is dominated by a single vortex without any possibility for the existence of mixing; therefore $U_0(H)$ reveals a weak magnetic field dependence. With the increasing magnetic field at a singular point, the pinning energy decreases and results in strong magnetic field dependence, this indicates the overlapping of vortices at high field region. The sharp decrease of $U_0(H)$ for $H > 2.5\,T$, confirms a crossover from a single to a collective flux pinning regime. As shown in figure 7(d), $\alpha$ is found to be $0.30 \leq \alpha \leq 0.22$ ($H < 2.5$ T) and $2.39 \leq \alpha \leq 2.35$ ($H > 2.5$ T) as pressure increases to 2.5 GPa. As pressure increases, $U_0(H)$ decreases, which implies that overlapping of both single and collective vortex states is realized in Re$_6$Hf.

The magnetic field dependence of critical current density, $J_c(H)$, is deduced from $M(H)$ curves for $T < T_c$ at different $P$. According to Bean's critical model [54,55], for a rectangular shaped sample, $J_c(H) = \frac{20\Delta M(H)}{l - (l^2 - 3b)}$, where $\Delta M(H)$ is the width of the superconducting isothermal magnetic loop and $l$ and $b$ are the length and width of the sample, respectively. The $J_c(H)$ at various temperatures are shown in figure 8(a) and (b) for pressure ranging from 0 to ~1.10 GPa. With increasing pressure, $J_c(H = 0, 2\,K)$ decreases from $1.1 \times 10^5$ A/cm$^2$ at 0 GPa to $0.64 \times 10^5$ A/cm$^2$ at 1.10 GPa. This decreasing trend of $J_c(H = 0, 2\,K)$ with pressure is similar to many Fe based and 2D layer [NbSe$_2$, (Fe$_{0.88}$Co$_{0.12}$)$_2$As$_2$] superconductors [56–58]. However, $J_c(H = 0, 2\,K)$ for Re$_6$Hf is several orders of magnitude higher than the high-$T_c$ cuprate superconductors $J_c(H = 0) \sim 10^4$-$10^5$ [59–61]. To obtain further insight into the nature of the vortex state in Re$_6$Hf, we analyzed the field dependence of $J_c$ using the collective vortex pinning model. Based on this model, $J_c(H)$ should follow a relation, $J_c(H) = J_0 exp^{\left(-\left(\frac{H}{H_0}\right)\right)^{\frac{3}{2}}}$, where $J_0$ and $H_0$ are the critical current density at 0 T and controlled parameter, respectively [62]. For all pressures (figure 8), $J_c(H)$ at different temperatures can be well described by the above relation, implying the collective pinning of vortices. For further information, we also analyzed the $P$ dependence of $J_c$ at 2 K at different $H$ as shown in figure 8(c). The linear fits to the data suggest that for all fields, $J_c$ decreases with increasing pressure at a similar rate $\frac{d(log J_c(H,T))}{dP} = 0.20 \pm 0.03 \frac{Acm^{-2}}{GPa}$, which could help one to understand the interaction between the vortices and the pinning center. Figure 8(c) shows $J_c(T)$ for different $H$ at 1.10 GPa. Also, information about the nature of the vortex pinning mechanism can be obtained by analyzing the temperature dependence of $J_c$ at a particular magnetic field. Based on Ginzburg-Landau theory, $J_c(T)$ at various $H$ can be fitted with a power law relation, $J_c(H) \propto \left(1 - \frac{T}{T_c}\right)^\beta$. The specific value of exponent $\beta$ signifies the vortex pinning mechanism in that magnetic field. For $\beta = \sim 1$ and $\leq 1.5$ indicate non-interacting vortices and core pinning mechanisms [63], respectively. From figure 8(d), it is clear that for $P$ up to 0.33 GPa, $\beta$ remains less than 1 in the low field region ($H < 0.15\ T - 0.25\ T$). However, with a gradual increase of field $\beta$ exceeds 1. Interestingly, irrespective of the field range $\beta$ is always greater than 1 for $P \geq 0.68$. More surprisingly, $\beta$ even exceeds $2$ at 1.1 GPa. These results demonstrate that collective pinning of vortices may be robust at high magnetic fields and high pressure regimes in Re$_6$Hf.

According to the collective pinning theory, the pinning of vortices arises from two processes. One is the pinning due to the spatial variation in mean free path, known as $\delta l$ pinning, and the other is due to the spatial variation of random distribution of $T_c$, called as $\delta T_c$ pinning. This theory predicts, $J_c(t)/J_c(0) \propto \left(\frac{(1-t^2)^{\frac{5}{2}}}{(1+t^2)^{\frac{1}{2}}}\right)$ for $\delta l$ pinning becomes, whereas for $\delta T_c$ pinning, $J_c(t)/J_c(0) \propto \left(\frac{(1-t^2)^{\frac{7}{6}}}{(1+t^2)^{\frac{5}{6}}}\right)$, where $t\ (=\ T/T_c)$ is reduced temperature. Figure 9 shows that the experimental data at 0 and 0.33 GPa fits well with the theoretical $\delta T_c$ pinning model at different magnetic field. This reveals that the $\delta T_c$ pinning is more dominant at ambient and lower applied $P$. At 0.68 GPa, the experimental data lies in between the $\delta T_c$ and $\delta l$ pinning curves, suggesting the coexistence of $\delta T_c + \delta l$ pinning. With a further increase of $P$ to 1.10 GPa, the pinning behavior changes completely to $\delta l$ pinning mechanism. A similar chemical and external $P$ induced change in flux pinning mechanism has also been reported on various superconductors such as MgB$_2$, NbSe$_2$, and pnictides respectively [57,64–67]. These results clearly show that $P$ can tune the pinning mechanism from $\delta T_c$ pinning to $\delta l$ pinning in Re$_6$Hf. Here, we would like to mention that the $\delta T_c$ and $\delta l$ pinning are directly proportional to coherence length $\xi$ and $1/\xi^2$, respectively. In fact, this observation of crossover pinning mechanism from $\delta T_c$ to $\delta l$ with increasing pressure is consistent with the increasing $\xi$ behavior estimated from $\rho(T, H, P)$ measurements (figure 6(c)).

Next, we study the magnetic field dependence of the flux pinning force, $F_p(H) = \mu_0 H \times J_c(H)$ at different pressures based on the Dew-Hughes model for a superconductor [68]. The scaling behavior of the pinning force density, $f(h) = Ah^m(1 - h)^n + Bh^p(1 - h)^q$, where $f(h) = \left(\frac{F_p}{F_p^{max}}\right)$ and $h = \left(\frac{H}{H_{irr}^k}\right)$ is a normalized pinning force density and reduced magnetic field and $(m, n)$, and $(p, q)$ are scaling constants used to describe the nature of the pinning mechanism of the sample. The $H_{irr}^k$ at various temperatures is estimated by linearly extrapolating the $J^{0.5}H^{0.25}$ $vs$ $H$ curve towards zero. From the analysis of Kramer's plot [69], the power exponent $(p, q)$ becomes (0.5, 2), (1, 2) and (0.5, 1) which indicates the surface boundary pinning, point pinning, and volume pinning, respectively. Normalized magnetic field $(h \sim H/H_{irr}^k)$ dependence of pinning force density $\left( f = \left( \frac{F_p}{F_p^{max}} \right) \right)$ at various temperatures with fixed applied $P$ is shown in figure 10. The Dew-Hughes model well fitted with the experimental data $(m, n)$ and $(p, q)$ is (0.76±0.04, 1.32±0.07) and (0.86±0.21, 2.46±0.46) respectively, at 2K reveals the coexistence of both volume and grain boundary pinning at ambient $P$. The extracted power exponents [$(m, n)$ and $(p, q)$] from the fitting analysis shows (0.87±0.04, 2.07±0.08) and (0.81±0.09, 2.60±0.06) at 0.33 GPa, (0.86±0.04, 2.38±0.09) and (0.78±0.09, 2.68±0.06) at 0.68 GPa and (0.86±0.04, 2.39±0.37) and (0.78±0.13, 3.10±0.09) at 1.10 GPa. With the application of $P$ (~0.33 GPa), the volume pinning is completely changed to point pinning, revealing that point pinning center is enhanced with the application of external $P$. Further, the surface pinning is more dominant than the point pinning above ~ 0.33 GPa possibly due to the relocation of grain boundaries in this polycrystalline sample.

We measured the pressure-dependent XRD pattern of Re₆Hf at room temperature using a Rigaku diffractometer to check for the possibility of any pressure-induced structural transition as shown in supplementary figure (S3). The powder diffraction pattern was processed by the CrysAlis pro (Rigaku oxford diffraction) software, and the crystal structure, lattice parameters are confirmed using the SHELXT and SHELXL-2018/3 methods. With pressurization the peak position move towards higher $2\theta$ angle suggesting shrinkage of the lattice parameters. In figure 11, we plot the calculated unit cell volume and lattice parameter. With pressure up to 18.01 GPa, the lattice parameters and unit cell volume are compressed by 1% and 3%, respectively. The smooth variation of these structural parameters rules out a structural transition up to 18.01 GPa. Furthermore, we did not detect any discernible new peak even at ~18 GPa which can signal the occurrence of a structural change. The bulk modulus $(B_0)$ is calculated from the unit cell volume using the first order Birch-Murnaghan equation of state [70,71], $P = \frac{B_0}{B_0'} \left[ \left( \frac{V_0}{V} \right)^{B_0'} - 1 \right]$. The large value of high $B_0 = 333.63$ $GPa$ suggests that Re$_6$Hf exhibits poor compressibility. This may be one reason for the modest change of $T_c$ with pressure.

The calculated electronic band structure using density functional theory is shown in figure 12(a-b). The total density of states (DOS) at Fermi level, $E_F$, together with the site projected partial DOS are shown in figure 12(c). In both ambient and high-pressure phases, the bands are crossover the $E_F$ and the DOS is small but finite, underlining the Re$_6$Hf to be metallic. The total DOS at the equilibrium volumes as a function of $P$ is displayed in figure 12(d). On a closure look to figure 12(e), DOS decreases barely up to 10.0 GPa and is followed by a slight decrease thereafter. This observation agrees well with the experimental results of a modest $T_c$ decrease up to 8 GPa. The DOS near the Fermi level is entirely composed of Re and $Hf - d$ bands, however, since the ratio of Re to Hf is 6:1, $Re - d$ bands comprise the majority of the states. In the entire valance band region, in addition to the Hf- and $Re - d$ states $Hf - p$ and $s$ states are predominant compare with the $Re - s$ and $p$ states.

**Conclusion**

In summary, we have investigated the superconducting and structural properties of a polycrystalline non-centrosymmetric superconductor, Re$_6$Hf. We found that as pressure increases to 8.0 GPa, $T_c$ decreases modestly, and the superconducting transition width becomes sharper. Analysis of the structure using XRD under pressure rules out any structural transition and a high value of bulk modulus $B_0 = 333.63\ GPa$. However, density functional theory calculations suggest a modest reduction of DOS at Fermi level with pressure, which may be a possible reason for such a small decrease of $T_c$ with pressure. Furthermore, both $H_{c2}(0)$ and $H_{c1}(0)$ decrease slightly from the ambient pressure value as pressure increases to 2.5 GPa. In addition, analysis of $\rho(T, H)$ using TAFF of vortices revealed a double linearity field dependence of activation energy of vortices $U_0(H)$, confirming the coexistence of single and collective pinning vortex states. Consistently, analysis of critical current density using the collective pinning theory showed the transformation of $\delta T_c$ to $\delta l$ pinning as pressure increases, possibly due to migration of grain boundaries.

**Acknowledgement**

S.M. thank to SERB for the financial support to visit ISSP, The University of Tokyo, Japan, through OVDF and acknowledges to ISSP for providing the experimental facilities during the visit and thanks to Dr. S. Nagasaki, ISSP, The University of Tokyo, Japan, for her continuous support for cryogenic measurements. The author SA acknowledges to SERB, DST (FIST), TANSCHE, UGC-DAE-CSR, Indore, DAE-BRNS and RUSA 2.0. Author YU acknowledges to JSPS KAKENHI Grants No. JP19H00648. MK thanks to UGC-RGNF for the meritorious research fellowship. R.P.S. acknowledges the Science and Engineering Research Board, Government of India, for Core Research Grant No. CRG/2019/001028. PV gratefully acknowledges the Research Council of Norway for providing the computer time (under the project number NN2875k and NS2875k) at the Norwegian supercomputer facility.

## References:-

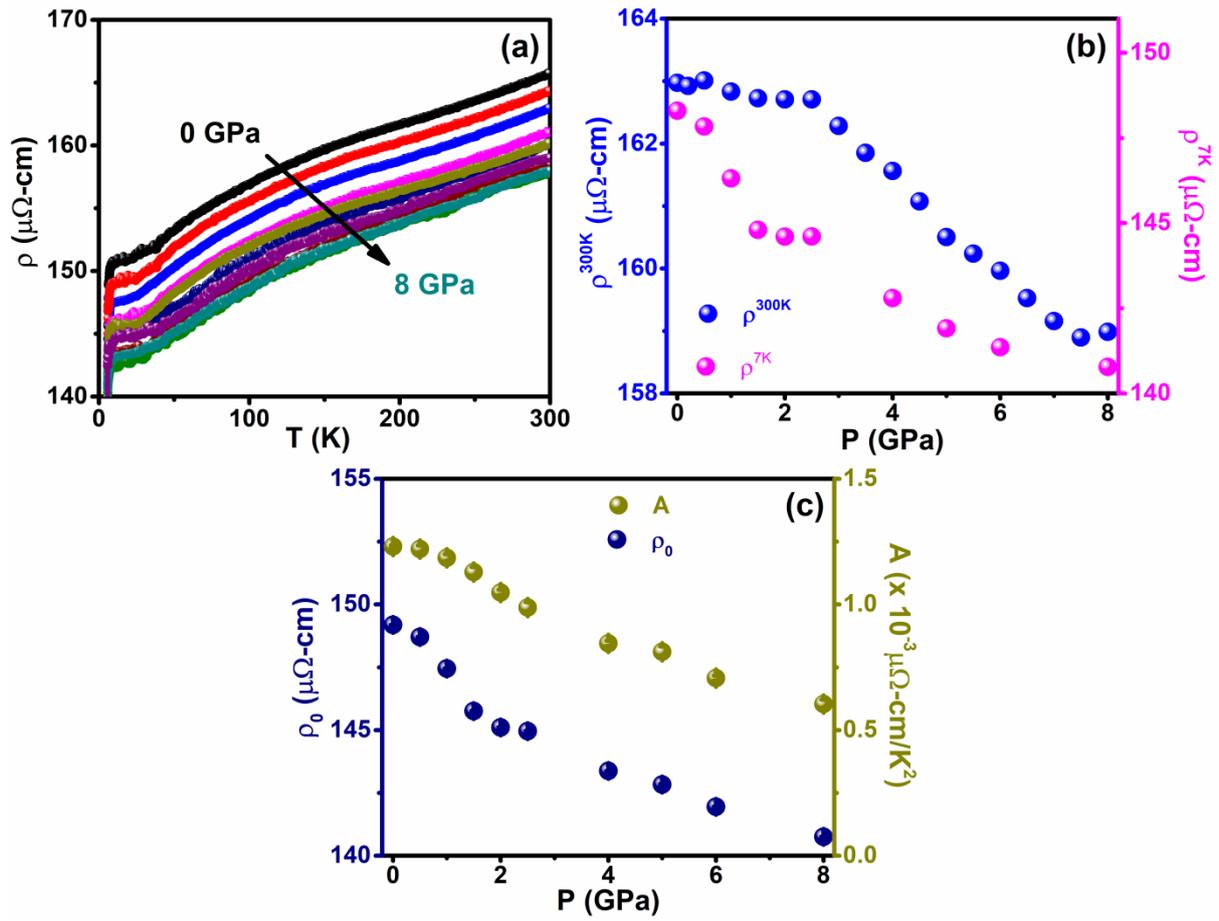

Figure 1 (a) Normal state $\rho(T)$ under various applied hydrostatic pressure, (b) $P$ dependence of resistivity at room temperature and 7 K ($\rho_{300\,K}$ and $\rho_{7\,K}$). (c) $P$ dependence of $\rho_0$) and scattering factor ($A$) determined from Fermi-liquid model fitting $\rho = \rho_0 + AT^2$.

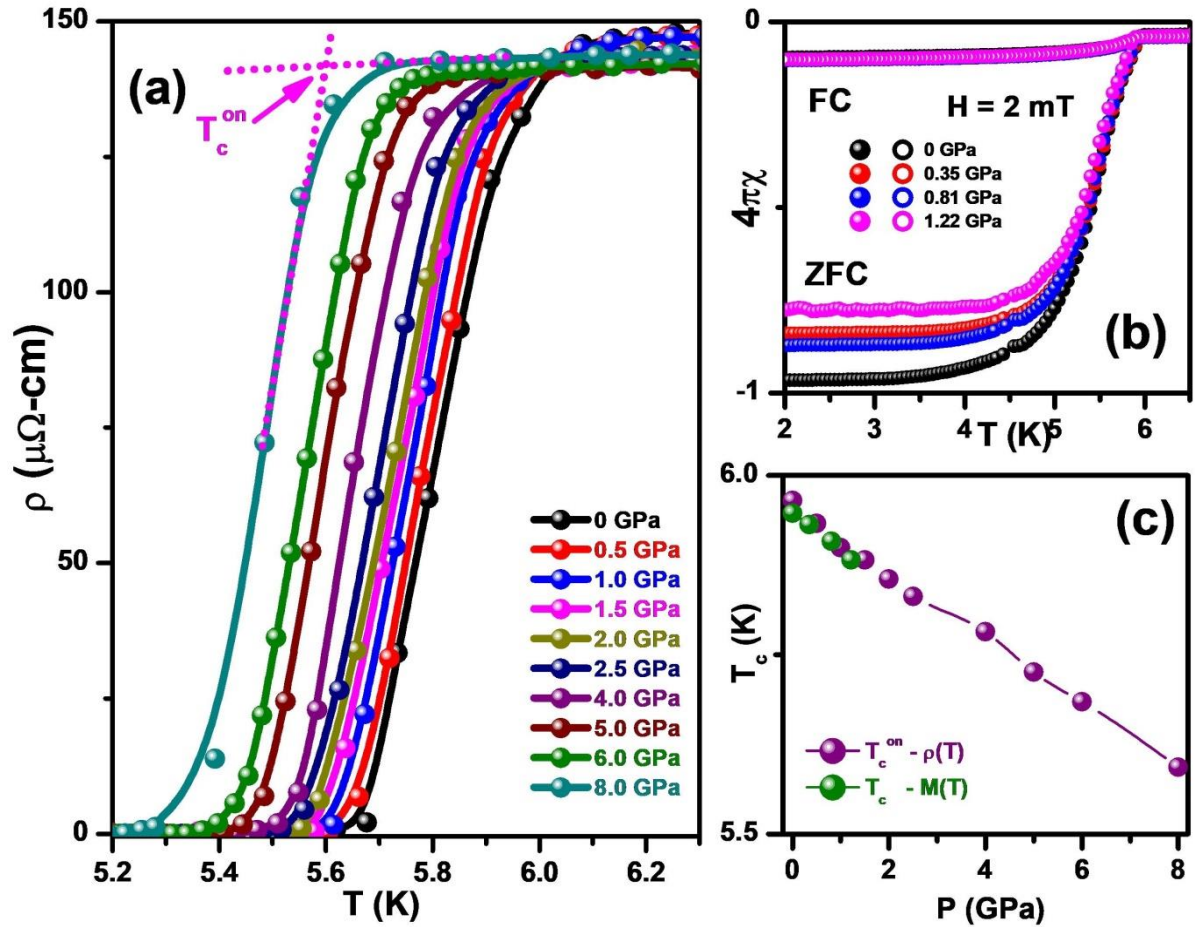

Figure 2 (a) $\rho(T)$ under various applied hydrostatic pressure near superconducting $T_c$ region, (b) Temperature dependence of unit less volume susceptibility ($\chi(T)$) with the magnetic field of 2 mT for $Re_6Hf$ and (c) $P$ dependence of onset of superconducting $T_c$ from $\rho(T)$ and $\chi(T)$ measurements for $Re_6Hf$.

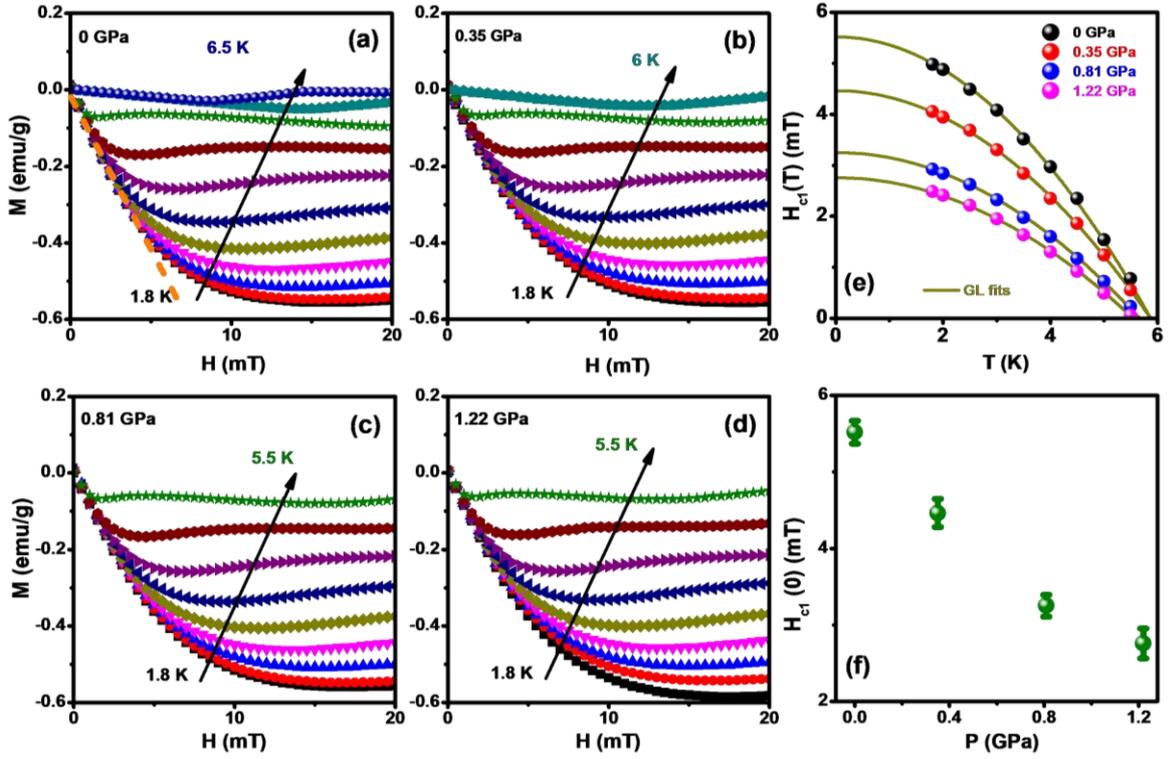

Figure 3 (a)-(d) $M(H)$ at various temperatures under constant $P$ of 0 GPa, 0.35 GPa, 0.81 GPa, and 1.22 GPa, (e) $H_{c1}(T)$ at fixed $P$ and the solid lines represents the G-L relation and (f) $P$ dependence of $H_{c1}(0)$ estimated from the G-L relation fits for $Re_6Hf$.

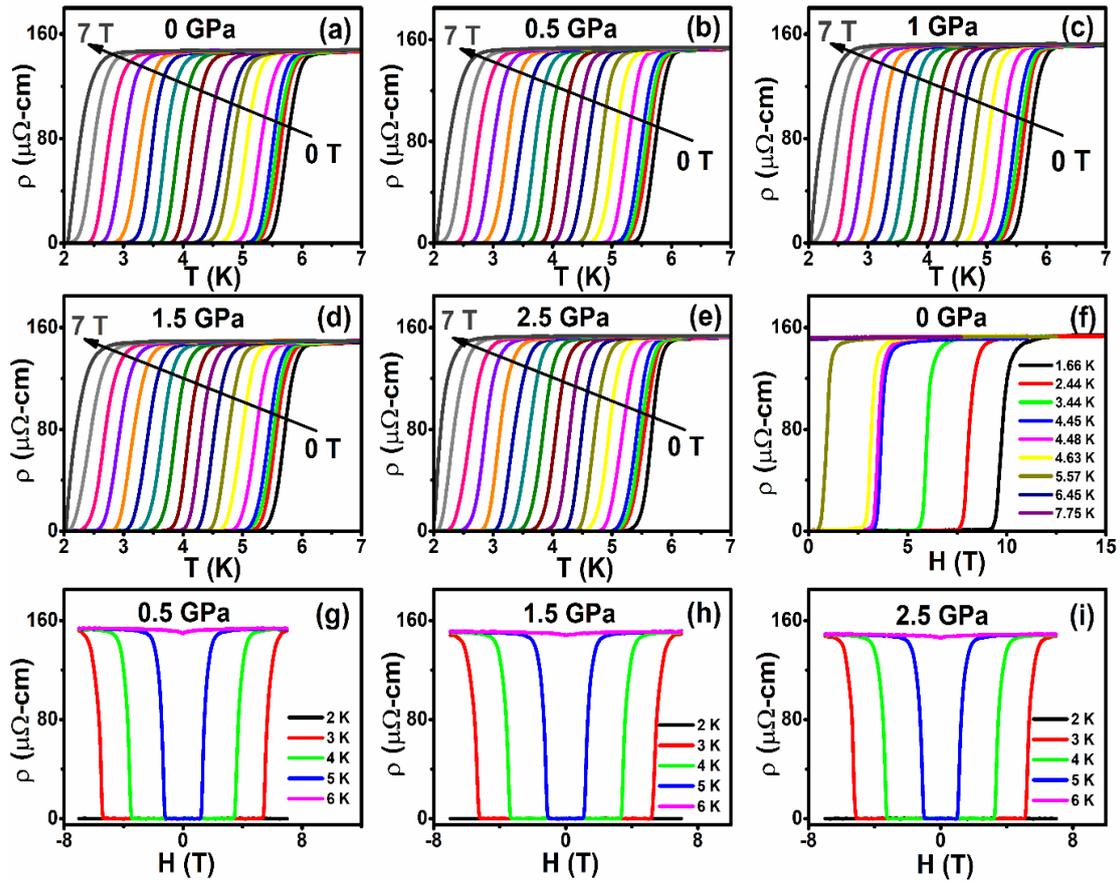

Figure 4 (a)-(e) $\rho(T)$ at various applied magnetic field under constant $P$ from 0 to 2.5 GPa and (f)-(i) $\rho(H)$ at various temperature with applied $P$ of 0 GPa, 0.5 GPa, 1.5 GPa and 2.5 GPa for $Re_6Hf$.

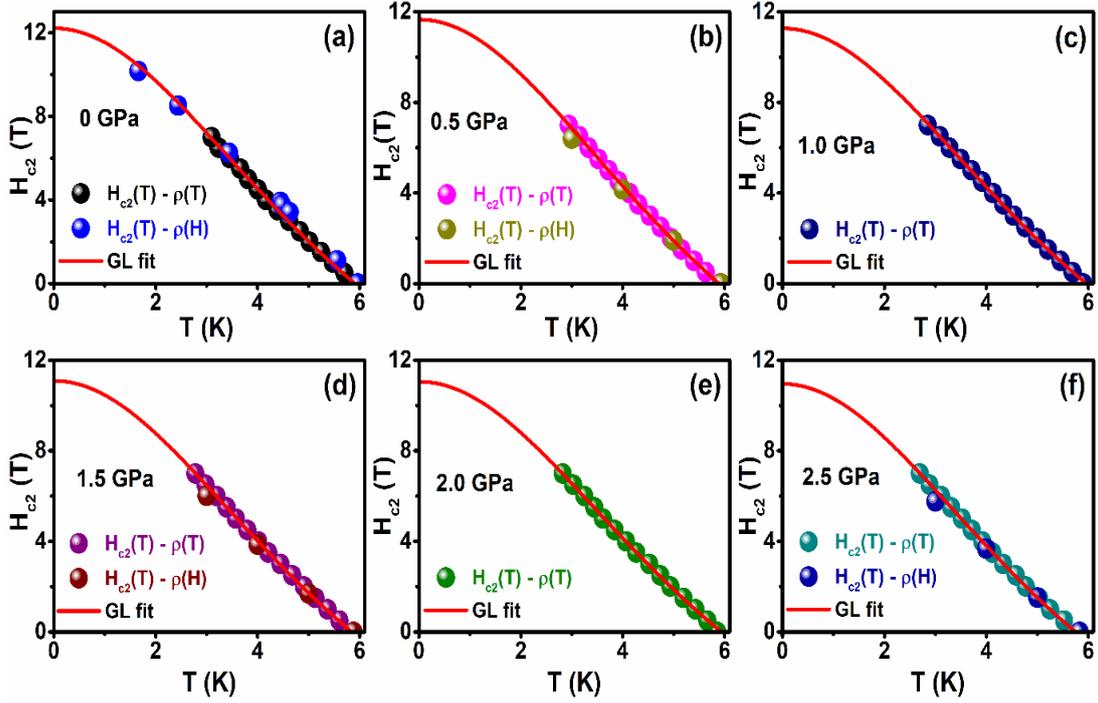

Figure 5 (a)-(f) $H_{c2}(T)$ determined from $\rho(T, H)$ measurements at various applied $P$ of 0 GPa, 0.5 GPa, 1 GPa, 1.5 GPa, 2 GPa and 2.5 GPa and solid lines are fitted with G-L relation yields $H_{c2}(0)$ for Re$_6$Hf.

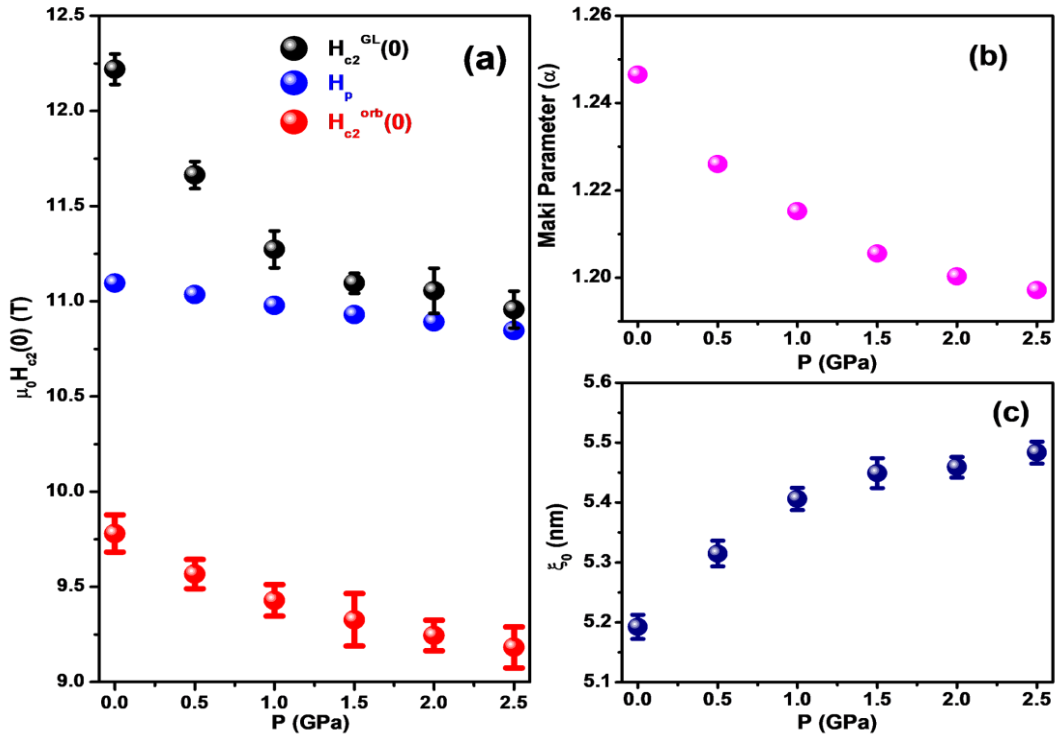

Figure 6 (a) $P$ dependence of upper critical fields ($\mu_0 H_{c2}^{GL}(0)$ & $\mu_0 H_{c2}^{orb}(0)$), and Pauli limited critical field ($\mu_0 H_p(0)$), (b) Maki parameter ($\alpha$) and (c) coherence length ($\xi(0)$) for Re$_6$Hf.

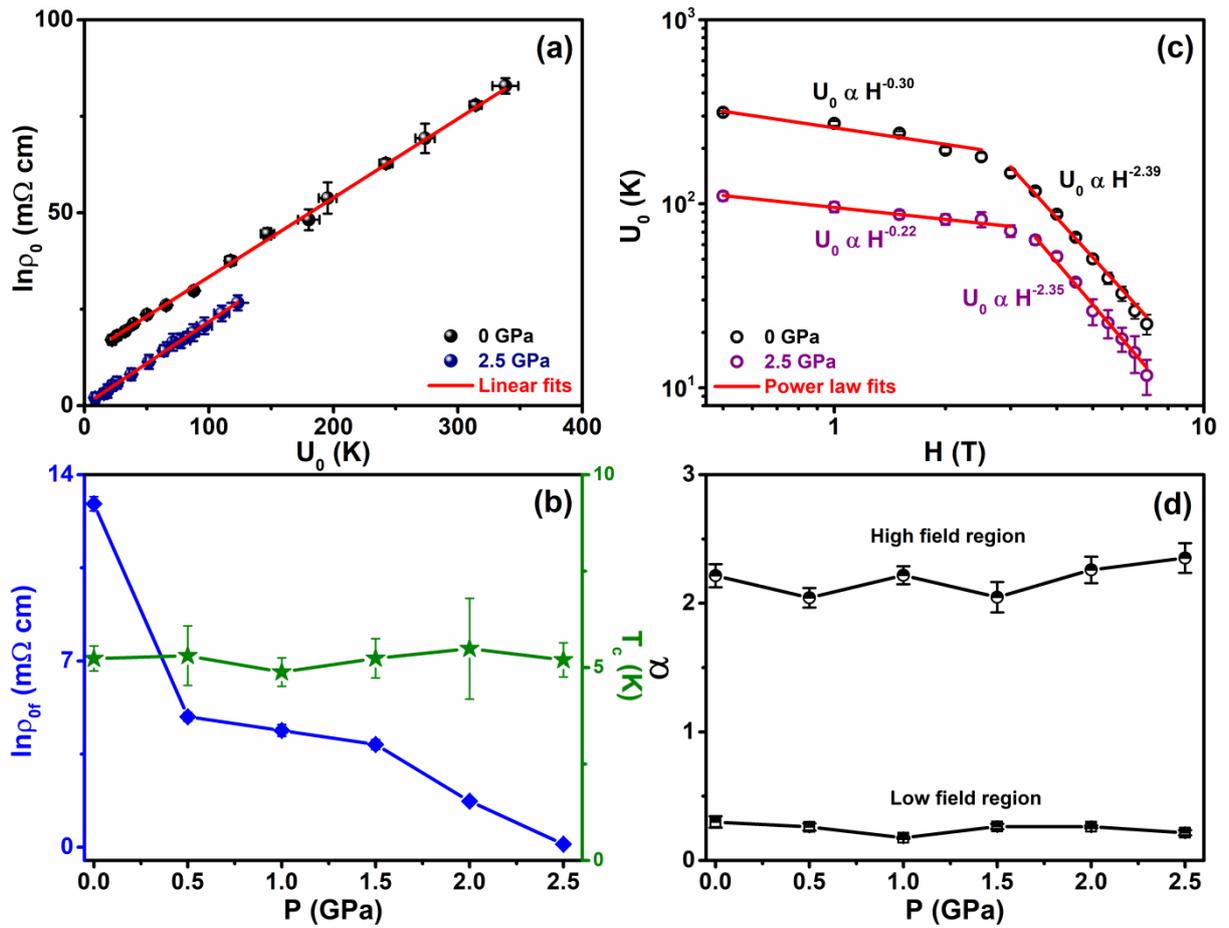

Figure 7 (a) $ln\,\rho_0(H)\,vs\,U_0(K)$ determined from the Arrhenius graph ($log\,\rho(T,H)\,vs\,1/T$ in different magnetic fields at $P = 0$ and 2.5 GPa for Re$_6$Hf. (b) Shows the $P$ dependent $\ln\rho_{0f}$ and calculated $T_c$. (c) $U_0\,vs\,H$ determined from the Arrhenius plot, $log\,\rho(T)\,vs\,1/T$, in different magnetic fields at $P = 0$ and 2.5 GPa for Re$_6$Hf. (d) Pressure dependence $\alpha$ value calculated from power law dependence of $U_0(H) \propto H^{-\alpha}$ at low field and high field region of $U_0$.

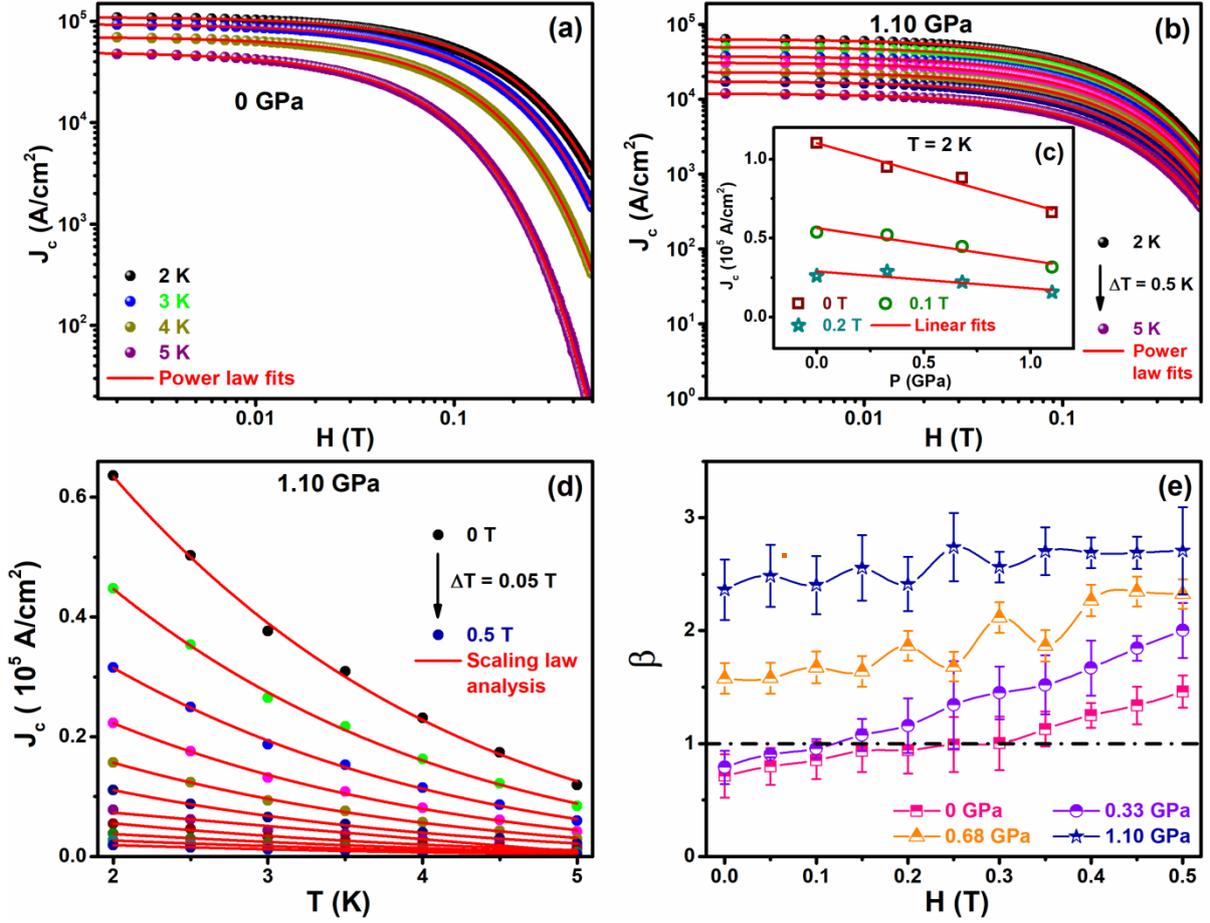

Figure 8: (a) and (b) shows $J_c(H)$ at different temperatures for $P = 0$ and 1.10 GPa, respectively, for Re$_6$Hf. Solid lines represent the power law behavior of the collective pinning model as described in the text. (c) $P$ dependence of $J_c(H)$ at 2 K and solid lines represent the linear fits to the data. (d) Temperature dependence of critical current density $J_c$ at different fields with $P = 1.10$ GPa for Re$_6$Hf. Solid lines represent the scaling law ($J_c(H) \propto \left(1 - \frac{T}{T_c}\right)^{\beta}$). (e) Magnetic field dependence of $\beta$ at different $P$. Dash dot line at $\beta = 1$ separates the single and collective vortex region.

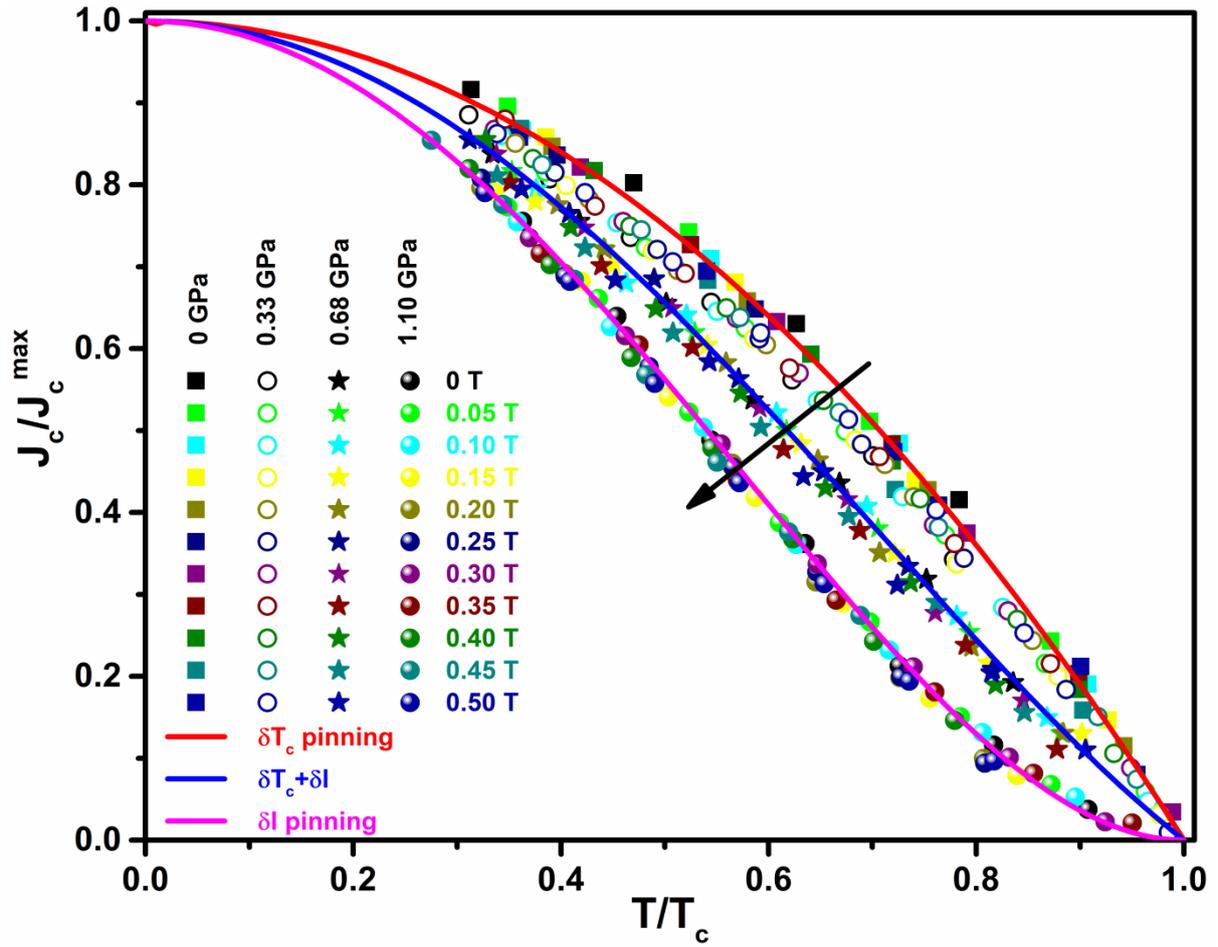

Figure 9 Normalized temperature dependence of $J_c$ at different magnetic fields with various $P = 0$, 0.33 0.68, and 1.10 GPa. Solid lines describe the fits to the data for $\delta T_c$, $\delta T_c + \delta l$, and $\delta l$-pinning based on the model of collective vortex pinning.

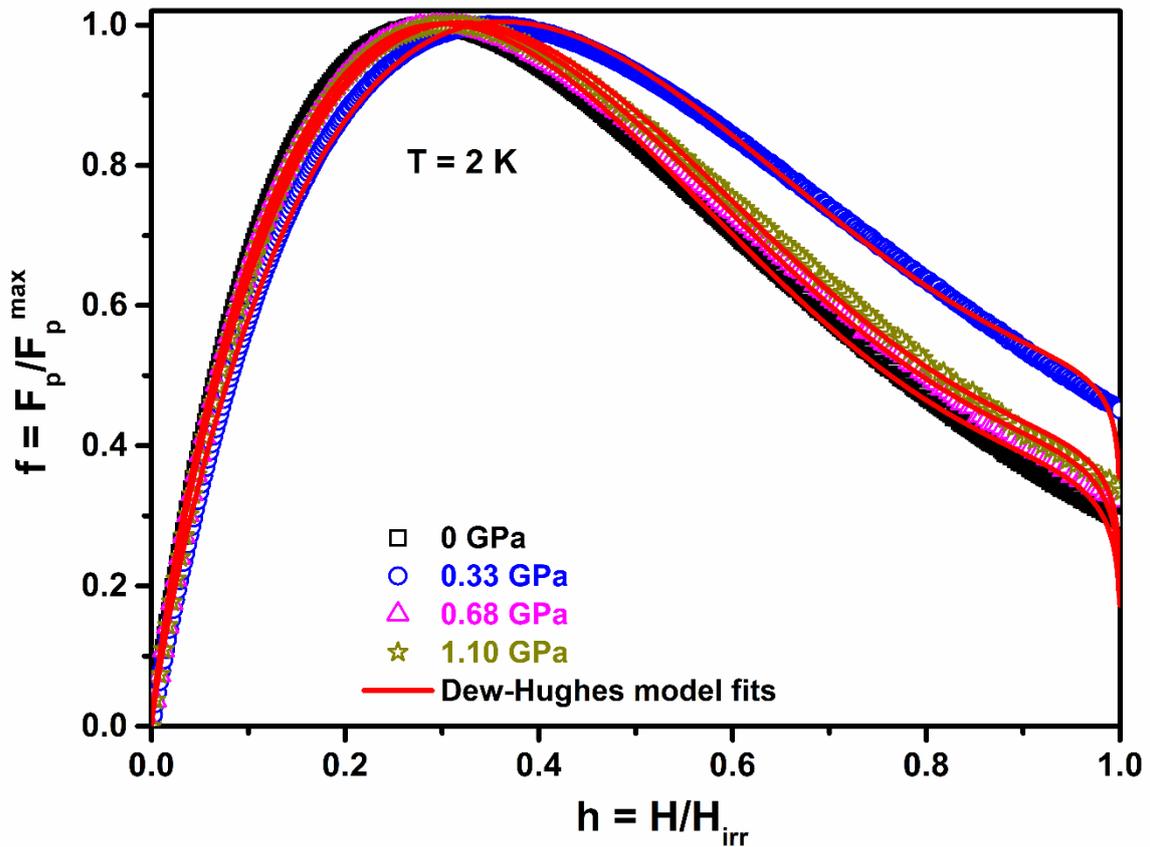

Figure 10 Normalized magnetic field dependence of normalized pinning force at 2 K with various applied *P* and solid lines represent the fitting of the Dew-Hughes model to describe the nature of flux pinning.

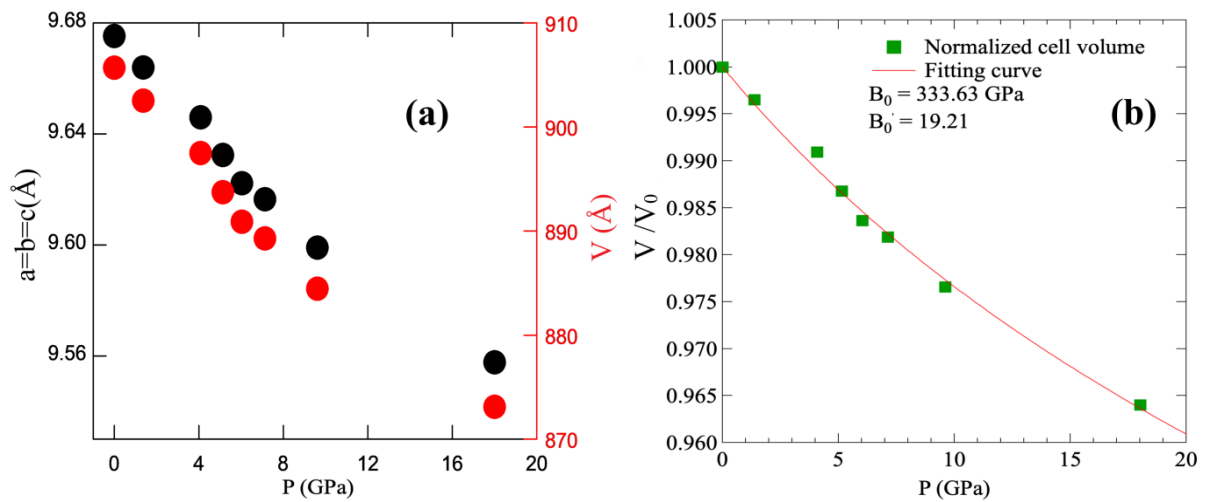

Figure 11. (a) Shows the pressure dependence of lattice parameters and unit cell volume. (b) Normalized volume ($V/V_0$) of the unit cell under *P*. The red solid line represents the Birch-Murnaghan equation of state fit to the data.

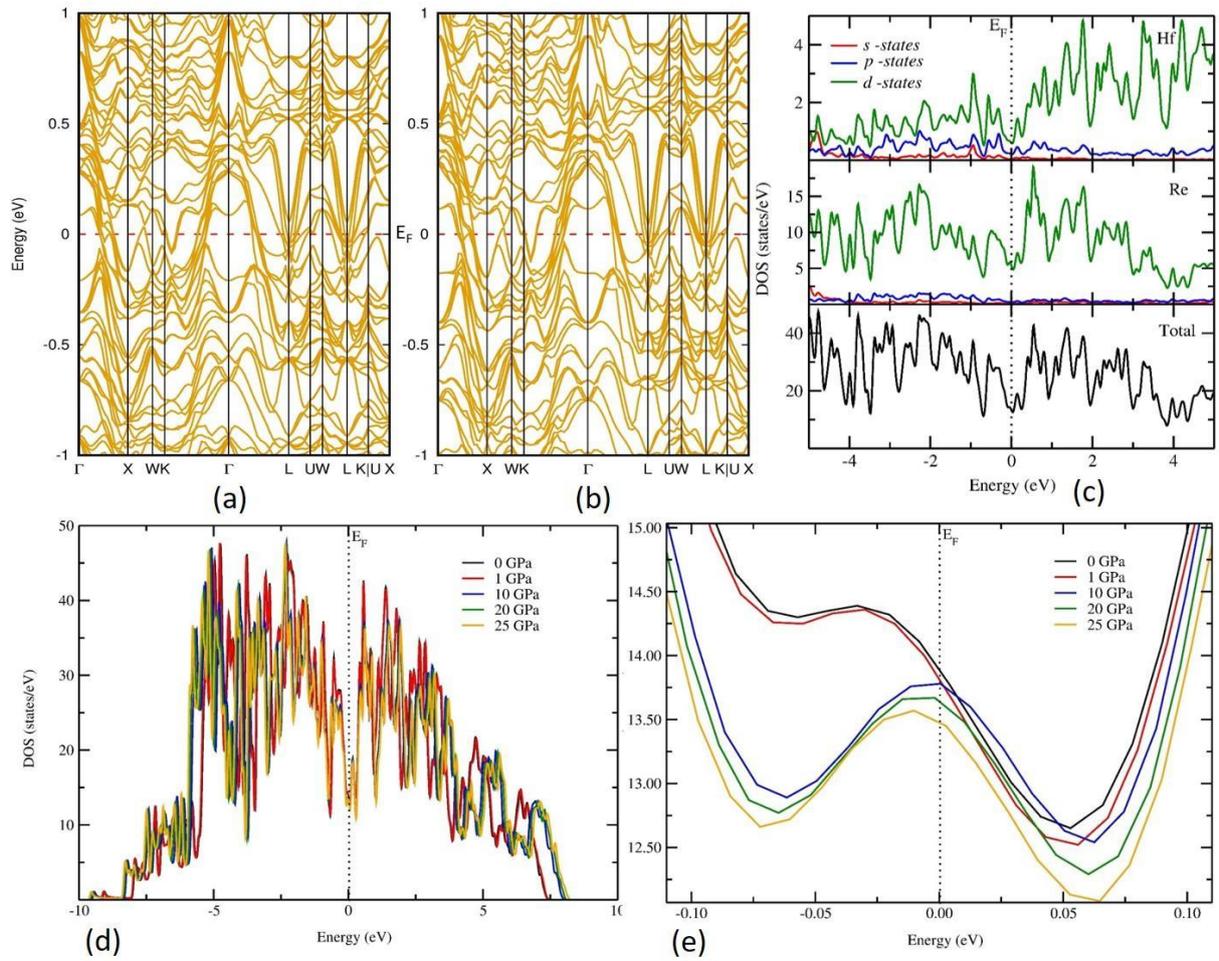

Figure 12 Calculated electronic band structure for Re₆Hf at ambient pressure [(a); 0 GPa)] and at 10 GPa (b). Total electronic density of states (DOS) and site and orbital projected DOS (c); total electronic DOS as a function of pressure (d); and the magnitude version at the Fermi level ($E_F$) is shown in (e) for Re₆Hf. GGA and PW91 functional are used for the computation. The Fermi level ($E_F$) is set to zero.